\documentclass[11pt]{article}
\begin{document}
%

\begin{center}
{\large \bf We live in the quantum 4-dimensional Minkowski space-time.}

\vskip.5cm


W-Y. Pauchy Hwang\footnote{E-mail: wyhwang@phys.ntu.edu.tw The contents
are originally for the introduction for the 2nd Edition of
"Relativistic Quantum Mechanics and Quantum Fields" (World Scientific,
2016) authored by W-Y. Pauchy Hwang and Ta-You Wu.}\\
{\em Department of Physics, National Taiwan University, \\
Taipei 106, Taiwan}

\vskip 0.2cm

{\small(April 17, 2015)}
\end{center}

\begin{abstract}
We try to define "our world" by stating that "we live in the quantum
4-dimensional Minkowski space-time with the force-fields gauge group
$SU_c(3) \times SU_L(2) \times U(1) \times SU_f(3)$ built-in from
the outset".

We begin by explaining what "space" and "time" are meaning for us - the
4-dimensional Minkowski space-time, then proceeding to the quantum
4-dimensional Minkowski space-time.

In our world, there are fields, or, point-like particles.
Particle physics is described by the so-called Standard Model.
Maybe I should explain why, how, and
what my Standard Model would be everybody's "Standard Model" some day.
Following the thinking underlying the minimal Standard Model and based on
the gauge group $SU_c(3) \times SU_L(2) \times U(1) \times SU_f(3)$, the
extension, which is rather unique, derives from the family concept that
there are three generations of quarks, on (123), and of leptons, on
another (123). It yields neutrino oscillations in a natural manner.
It also predicts a variety of lepton-flavor-violating rare decays.

At the end of the Standard Model, we will provide some clear
answers toward two "origin" questions: What is the origin
of mass? Another one: what is the origin of fields (point-like
particles)?

\bigskip

{\parindent=0pt PACS Indices: 12.60.-i (Models beyond the standard
model); 98.80.Bp (Origin and formation of the Universe); 12.10.-g
(Unified field theories and models).}
\end{abstract}

\bigskip

\section{The Space and Time}

I assume that you know about the length measurement and the standard clock. Here
we assume that the operations of sticks and clocks dees not affect anything.
Further. we can make sure that the local space-time which we are working with is
flat. If we label an event or point as $(x,y,z, ct)$ or simply $(x,y,z,t)$, then
the interval between two points, $(x_1,y_1,z_1,ct_1)$ and $(x_2,y_2,z_2,ct_2)$,
would be an invariant quantity, according to relativity. Thus, we, or you and I,
could begin to talk about something and in fact set up some common units for space
and time.

So, we have, with $\Delta x = x_2 - x_1$, $\Delta y= y_2 - y_1$, $\Delta z =
z_2 -z_1$, and $\Delta t = t_2 - t_1$,
\begin{equation}
(\Delta s)^2 = (\Delta x)^2 + (\Delta y)^2 +(\Delta z)^2 - (c\Delta t)^2,
\end{equation}
an invariant quantity among all the observers. Or, in infinitesimal forms,
\begin{equation}
ds^2 = dx^2 + dy^2 + dz^2 - c^2 dt^2 \equiv g_{\mu\nu} dx^\mu dx^\nu
=\eta_{\mu\nu} dx^\mu dx^\nu.
\end{equation}
Here $ds$ is called "the proper length", or "the proper distance",
$g_{\mu\nu}$ is the second-rank metric tensor, and
$\eta_{\mu\nu}=diag (1,1,1,-1)$. Similarly,
the $d \tau$ (with $d\tau^2 \equiv - ds^2$) is "the proper time". We shall
use "the invariant distance" or "the invariant time" for the sake of
being precise.

From the measurements of the space and of the time which are rather
independent, there is no reason why the Minkowski constraint, the equation
specified above, should be valid. Thus, we live in the 4-dimensional
Minkowski space-time.

\medskip

{\it Let's come to think about it. The so-called "intuition" (classically)
could be completely wrong. Because of "Lorentz contraction" and "time
dilation" already built in the above Minkowski's metric, the world of
an elementary particle differs from our limited one, based on classical
physics or Newton's. And, for this book on "Relativistic Quantum
Mechanics and Quantum Fields" \cite{Book}, we suggest that we should
imagine that our "intuition" of the world be the same as that for an
elementary particle. Thus, we try to downplay the importance of the
non-relativistic descriptions.}
\bigskip

{\it Moreover, the quantities $\Delta x$, $\Delta y$, $\Delta z$,
and $\Delta t$ are "observables" in the quantum mechanics sense.
Similarly, we should classify $\Delta s$ as a quantum-mechanical
observable.}

{\it These quantities should be realized as the quantum mechanical
observables, as to be indicated by the statements in the next
section. In other words, we start out by viewing these quantities
as the operators, not simply by commuting real numbers. This is the
case, even though in certain representations part of them are real
numbers.}

As another important example of the 21st-century post-modern physics,
let's start with the Schwarzschild metric,
\begin{equation}
d^2s=-(1-{2GM\over r})dt^2 + (1-{2GM\over r})^{-1} dr^2 +
r^2(d\theta^2+sin^2\theta d\phi^2).
\end{equation}
We should try to establish
our intuitive explanation of the time and the space, at all $r$ and
all $t$. Sign switch may be identified as the important source of
confusion. In this example, we use the natural units. We should
eventually establish a decent 21st-century physics, with this
"black hole" solution as one of the beginning points.

To this end, we identify $t$ as our time and $(x_1, x_2, x_3)$
as our space. The problem with the Schwarzschild metric is the
coefficient of $dr^2$ which goes over to $\infty$ at the horizon,
i.e., ${2GM \over r} \to 1$. The coefficient of $dt^2$ would go
to zero simultaneously. So, in our space and our time, maybe we
never get to this point - that is, black holes in our space-time
never form and all the black holes are pre-formed. Einstein equation
tells us so.

The other way of saying it: We try to write the Schwarzschild
metric as follows:
\begin{equation}
(ds)^2 = - c^2 (d t^*)^2 + (d r^*)^2 + r^2(d^2\theta + sin^2\theta d\phi^2).
\end{equation}
Here we have
\begin{equation}
{d r^*\over dr} = \pm \sqrt {r \over r-r_h}, \quad
{dt^*\over dt} = \pm \sqrt {r-r_h \over r}; \qquad r_h= 2G M.
\end{equation}
So, it becomes pure imaginary beyond the horizon ($r=r_h$) rather
than the switching between time and space, which might be wrong
at the first place. Our space variables and our time variable
are real numbers, i.e., $(-\infty,\, +\infty)$, after all. This
may offer an entirely new road for the knowledge of black holes.

We thus suggest that we may stick to the interpretation of
viewing $t$ as the time and $(x_1, x_2, x_3)$ as the space.
Thus, these are quantum mechanical observables.
Let these variables run from $-\infty$ to $+\infty$ to define
the observable space-time for us. We should not talk about the
events beyond us - not within the $\pm \infty$. We will be
"conservative" in this regard.

\medskip

We may choose the $cgs$ unit system, the length in $cm$, the time in
$sec$, and the mass in $gm$, so that the light velocity $c = 3.00
\times 10^{10} \,cm/sec$. Or, the natural units may be chosen such that
$\hbar = c=1$. Unless specified otherwise, the natural units will be used
throughout this textbook.

\medskip

We have to look at the space and the time in particular not too
"ideologically". First of all, the space-time changes according
to the matter distribution - if we adopt Einstein's general relativity.
In other words, the "definition" of the space-time varies with the matter
distribution, which could be anything. Then, how do we "define" the
matter in this context?

For example, in the Schwarzschild metric given above which is the
solution to Einstein general-relativity equation of a "point" mass,
the rationale of the "black-hole" solution still stirs up so many
puzzles, even until today (for about a hundred years later). So,
the mix-up of the "simple" space-time notion with the matter
notion sort of drives these concepts with a lot of ambiguities.

In fact, we might regard the "space-time" as a whole as the "physical
system" of some kind. In the simplest case, the physical system possesses
the Lorentz invariance and others. When we talk about a "point-like"
particle in the space-time, we could try to "define" what the "point-like"
means in the physical system of the space-time. In other words, the
space-time is more physical than mathematical.

In the last part of the book, we shall study the Standard Model
of particle physics. Electrons, other leptons, quarks, etc.,
the so-called "building block of matter", all
are point-like particles - up to the size of $10^{-20}\, cm$
(the known resolutions), these particles are still point-like.

So, we are studying the behaviors of these "points" in our
space-time, physically rather than mathematically, at the scale
far small than the atomic size of $10^{-8}$ cm.

\bigskip

\section{The Point in the Quantum Sense}

\medskip

\parindent=12pt

Mathematically, a "point" does not have any size or volume, according to the
geometry. When we talk about a "point-like" particle in the physical sense; it
should be different from the "point" in the mathematical sense. When
we describe a particle by Dirac equation or by Klein-Gordon
equation, that does not have the size parameter, we call it "point-like",
or "point-like-ness" in the quantum sense.

Quantum mechanics, or more precisely quantum principle, comes to rescue
in a mysterious way. At the scales around $1\, cm$, we deal with the system
macroscopically and we developed the classical physics. At the scales of
about $10^{-8}\,cm$, we know that quantum principle has to be there. The
uncertainty principle and others are working there, down to the scale of
$1\,fm$, or $10^{-13} cm$, or much smaller.

The quantum principle states that the position $(x, y, z)$ does not
commute with the momentum $m {d\over dt} (x,y,z)$, with the commutator
equaling $i \hbar$. Thus, we have to treat the position $(x,y,z)$ and
the momentum $m {d\over dt} (x, y,z)$ as operators when we speak of them
simultaneously. Thus, we have
\begin{equation}
[x_j, p_j]=i\hbar \delta_{ij},\qquad [x_i,x_j]=0,\qquad [p_i,p_j]=0.
\end{equation}
$\hbar$ sets the scale when the quantum effects are
"visible".

So, what is the concept of "point" in this quantum regime? It is relevant
for the atomic size (i.e. $10^{-8} cm$) or for the size of a nucleus
(i.e. $10^{-13} cm$). Eventually, it has become
rather murky at the distance of $10^{-20}cm$ or shorter. $10^{-20} cm$ is
somewhat smaller than the resolution set by the Large Hadron Collider
(LHC) at Geneva.

\medskip

We should mention the following:

When it becomes much smaller, such as $10^{-25} cm$ or smaller,
which are so small that we never get there,
the coordinates $(x,y,z,ct)$ might not commute among themselves --
the so-called "non-commutative geometry". This phenomenon, if it
exists, which we call it the "super-quantum regime" -- it is
rather natural in the line of "reasoning".

This noncommutative aspect among the coordinates was raised by Hartland
S. Snyder in as early as 1947 [Phys. Rev. {\bf 71}, 38 (1947)]
\cite{Snyder}. This aspect was discussed later on in a variety of
contexts. The idea could be relevant when we discuss the alternatives
of "the point-like structure" or point-like particles.

In any event, it is quite clear that the "point-like" in the physical sense could
be so different from the "point" in the mathematical sense. As a
physicist, we'd better keep in mind these subtleties, particularly when
we "polish" our theories to eventually describe more peculiar systems or
objects.

\bigskip

\section{Our World: The quantum 4-dimensional Minkowski space-time}

Thus, we propose that the differences of the space-time coordinates
are the quantum observables which are subject to the measurements
for realizing their values. The measurement of the space-time
coordinates is the very beginning of everything. As the quantum
law requires, the spatial coordinates do not commute with
the time derivatives of the spatial coordinates themselves -
implying that they are operators, or mappings or functions.

Based on this proposal, the inside of the black hole requires
the coordinates to be pure imaginary, thus beyond the
observability. Einstein equation relates Ricci's tensor
to the energy-momentum tensor, so mathematically it could
cover the "un-physical" region(s).

In other words, the laws of gravity, when it hits the
boundaries of the infinities or of the pure imaginary,
remain to be completely open. This is one of the
frontiers that we should pursue after, in this 21st
century.

{\it To emphasize, we live in the quantum 4-dimensional Minkowski
space-time. All the classical notions are in fact used to illustrate
our real space-time, nothing any more. In the following, we shall
mention that this space-time is also endorsed with the force-fields
gauge fields, according to the Standard Model.}

\bigskip

\section{The Standard Model of the 20th Century}

\bigskip

\parindent=12pt

Quantum mechanics and Einstein's theory of special relativity constitute the
basis for the main body of modern physics developed during the first half of
the twentieth century (that is, the two pillars of the 20th-Century modern
physics). Maybe, in the 21st-century physics, we should, carefully, introduce
the third pillar, the fourth pillar, and so on.

Physical systems that can be described by relativistic wave equations include
electrons, quarks, and many other elementary particles in the subatomic world.
The presentation on the Klein-Gordon equation could be simple since it has the
same symmetry as the background, i.e., the 4-dimensional Minkowski space-time.
It describes spinless particles such as the Higgs particles, for which the
example for its existence was in 2012, and the pions, the composite systems
which were discovered in 1940's. So after we explain
the axiom box for the meaning of "quantization" on the language, we shall
present the Dirac relativistic
equation in great details. The Dirac theory for the electron has many
successes:  It is Lorentz covariant; it contains the electron spin with the
gyromagnetic ratio $g = 2$; it predicts the anti-particle positron which is
experimentally discovered about ten years later; it gives the correct fine
structure of the hydrogenic atom levels; etc.  Application of the Dirac
equation to other leptons such as muons and neutrinos, and to quarks in
the context of bag models, also leads to quantitative successes.  It is
clear that, unlike atomic or molecular physics where relativistic
effects can often be treated as small perturbations, a suitable
introduction to elementary particle physics and field theories must commence
with a description of relativistic wave equations.

Early on, there were "difficulties" too with the Dirac equation.  One was
that of extending the theory for a single electron to a system of many
electrons.  Another was of an even more basic nature, namely, a theory
started out to represent a single electron ends up being inseparably
bound with a many-body effect on account of
the infinite sea of electrons in negative energy states.
In fact, with the discovery of the positron (the anti-particle of
the electron), the idea of the "negative-energy sea"
could have died off, but it persisted until the end
of the 20th Century.

But the antiparticle of the electron, or the positron, was discovered.
It explains why the Dirac theory must have four components - two for the
electron and another two for the positron.  The electron and the positron
are born to be described by the same Dirac equation. There is no such thing
which are called as " negative-energy states". The entire things could
have been clear from there, then. The notion of "negative-energy states"
should not appear, and in fact no need to appear --- we could call it
"the Mistake of the 20th Century Physics".

Historically, the story is as follows: The theory of quantized
electromagnetic fields begins with the work of Dirac in
1927, followed immediately by the work of Jordan and Wigner, and by Fermi in
1930.  A breakthrough comes only in the mid 1940's with the work of Tomonaga
in Japan and Schwinger, Feynman, and Dyson in the United States.  Although
infinities still remain, the theory succeeds in "subtracting" them away in a
definite, covariant way so that finite results can be obtained, which have been
found to be in excellent agreement with the observed Lamb shifts and the "$g-2$
anomaly".  The decade from the mid 1940's to the mid 1950's
is a period of fervent studies of quantum electrodynamics, both in further
calculations on this "renormalization" theory and in attempts to rid the theory
of the infinities (not just to isolate and bury them, so to speak).  Till now,
there seems not yet a satisfactory solution of this problem and in the
meantime physicists have turned their interest to other areas - elementary
particles and the nature of their interactions and their unification.\footnote{
\it Dirac believes that in a completely satisfactory theory, infinities should
not appear.  Since the late 1940's, he set out to re-examine and reformulate
classical dynamics and electromagnetic theory with a view to find a
different basic theory for quantization.  He believes even some new mathematics
not yet known may be needed.  He expressed his views only occasionally in
writing, but much more freely in private discussions. \smallskip}

During the decade from the mid 1950's to the mid 1960's, serious attempts were
made in searching for an alternative means of describing interactions among
elementary particles in terms of the so-called "S-matrix theory", in which one
tries to determine the scattering amplitudes, or the S-matrix elements, using
general principles such as unitarity and microscopic causality [which lead to
dispersion relations] and a minimal set of dynamical assumptions.  Altogether
in the S-matrix approach, the question concerning the underlying dynamics must
be answered, or postulated, before any quantitative predictions can be
consistently made.  Therefore, the notion of using "gauge field
theories"\footnote{\it QED is the simplest prototype of gauge field theories,
in which there exist
a class of so-called "gauge transformations" which do not affect physical
observables (classically). The notion of a gauge transformation will
be introduced later and it will become the main theme (idea) in the
second and third parts of this book.\smallskip} to
describe interactions among "building blocks" of mater has scored an amazingly
successful comeback since the late 1960's while progresses in the S-matrix
approach, which have since been relatively limited, have become things of the past,
at least for the time being.

The development of particle physics of the last half century [since the late
1960's till the present - 2015] consists of several major breakthroughs which
culminate in the general acceptance of the $SU(3)_{color} \times SU(2)_{weak}
 \times U(1)$ gauge field theory of strong, electromagnetic, and weak
interactions as the "standard model" \cite{PDG}. The Glashow-Salam-Weinberg [GSW]
$SU(2)_{weak} \times U(1)$ gauge theory provides a unified description of
electromagnetic and weak interactions.  It predicts the existence of neutral
weak interactions, the existence of the charm quark, and the existence of weak
bosons $W^\pm$ and $Z^0$ [which mediate weak forces], all of which have been
substantiated experimentally.\footnote{\it In a short time span of ten years
starting from 1976, Nobel prizes have
been awarded three times to discoveries related to the GSW electroweak theory:
for the discovery of the charm quark, for the successes of the GSW theory, and
for the experimental observation of $W^\pm$ and $Z^0$.} In the meantime, the
quantized $SU(3)_{color}$
gauge theory, or quantum chromodynamics [QCD], has been established as the
candidate theory of strong interactions among quarks and gluons, which are
believed to be the building blocks of all observed strongly interacting
elementary particles or hadrons.  QCD supports the dual picture of considering,
e.g., a proton as a collection of almost noninteracting quarks, antiquarks, and
gluons at high energies [because of the asymptotically free nature of QCD] and
as a system of confined, dressed valence quarks at low energies [because
QCD is consistent with color confinement].

What is troubling the theoretical physicists over almost a whole century is the
occurrence of ultraviolet divergences - see Chapter 10 for example. If we
"believe" our Standard  Model, the infinite parts should cancel out if
all ultraviolet divergences of the same characteristics all are taken account.
The answer to this question might be on the positive side.

In Part B of the present volume, we shall present in some detail quantum
electrodynamics, the simplest prototype of all quantized gauge field theories.
We also describe the conventional standard model, i.e., QCD and the GSW
electroweak theory, and the experimental tests of it. In Part C, we move on
to describe the Standard Model of the 21st century. We hope to conclude
the book with the real Standard Model, the real final chapter.

\bigskip

\section{Building Blocks of Matter}

\medskip

\parindent=12pt

We shall for convenient reference begin with a qualitative summary concerning
the subatomic and atomic world. In the minimal Standard Model, building
blocks of matter are known to include (a) three
generations of fermions, (b) mediators of fundamental interactions, and (c)
scalar particles which are responsible for spontaneous symmetry breaking
related to the physical vacuum.
Specifically, fermions consist of leptons, quarks, and their antiparticles:

\bigskip

\parindent=0pt
{\it Leptons:}

\begin{equation}
(e^-,\ \nu_e),\ \ (\mu^-,\ \nu_\mu),\ \ (\tau^-,\ \nu_\tau),\quad (columns).
\end{equation}

\parindent=0pt
{\it Quarks:}
\begin{equation}
\left(\begin{array}{ccc}
(u_R,\ d_R),& (c_R,\ s_R),& (t_R,\ b_R),\\
(u_Y,\ d_Y),& (c_Y,\ s_Y),& (t_Y,\ b_Y),\\
(u_B,\ d_B),& (c_B,\ s_B),& (t_B,\ b_B).\\
\end{array}\right) \quad (columns)
\end{equation}

\parindent=0pt
Note that leptons include electrons $(e^-)$, muons $(\mu^-)$, tau-leptons
$(\tau^-)$, electronlike neutrino $(\nu_e)$, and so on.  Quarks come in with
six possible flavors:  up ($u$), down ($d$), charm ($c$), strange ($s$),
bottom ($b$), and top ($t$).
A quark of any given flavor is assumed to carry one of three
possible colors:  red ($R$), yellow ($Y$), blue ($B$), or,
$x$, $y$, and $z$ with

\begin{equation}x=(1,\, 0,\, 0), \quad
y=(0,\, 1,\, 0), \quad
z=(0,\, 0,\, 1), \quad (columns).\end{equation}

\parindent=0pt
Quarks are not observed in isolation presumably because color is strictly
confined, a property consistent with the conjectured two-phase picture of QCD.

\parindent=12pt

Mediators of fundamental interactions include (1) the photon ($\gamma$), which
mediates the well-known electromagnetic interaction, (2) three weak bosons
($W^\pm, Z^0$), which mediate charged and neutral weak interactions, and (3)
eight gluons, which mediate strong interactions among quarks and antiquarks.
Gluons carry one of eight possible octet colors and cannot exist in isolation
because of color confinement.\par
{\it Hadrons}, or strongly interacting elementary particles, are by assumption
color-singlet, or colorless, composites of quarks, antiquarks, and gluons.
The hadrons include mesons, baryons, glueballs, and so on.  Pions ($\pi^\pm,
\pi^0$), kaons ($K^\pm$, $K^0$, ${\bar K}^0$), etas ($\eta, \eta'$), rho-mesons
($\rho^\pm, \rho^0$), $\psi/J$, and upsilons ($\Upsilon, \Upsilon',....$)
all are
{\it mesons} which, at low energies, are believed to be quark-antiquark pairs
confined to within the region defined by the meson size.  Nucleons (p, n),
lambda ($\Lambda$), sigmas ($\Sigma^\pm, \Sigma^0$), xi's ($\Xi^-, \Xi^0$),
deltas ($\Delta^-$, $\Delta^0$, $\Delta^+$, $\Delta^{++}$), charmed lambda
($\Lambda_c$), and bottomed lambda ($\Lambda_b$) all are {\it baryons}
which, at low
energies, look like systems of three quarks confined to within the region
defined by the baryon size.  {\it Glueballs} are colorless objects
consisting of gluons only.\par
{\it Nucleons}, i.e., protons (p) and neutrons (n), are believed to be primary
building blocks of the {\it nuclei} of the various atoms, ranging from the proton
itself [as the simplest nucleus], to the deuteron, $\alpha$, $^{12}C$,
$^{56}Fe$, $^{208}Pb,$ and
even to neutron stars [$A = \infty$ nuclei].  Replacement of a
nucleon in an ordinary nucleus by a lambda ($\Lambda$) or by a sigma-baryon
($\Sigma^\pm, \Sigma^0$) results in a {\it hypernucleus}.  An {\it atom}
is a system of
electrons around an ordinary nucleus, the whole system being often electrically
neutral.  {\it Molecules} are built from atoms.  All matter observed
terrestrially or
celestially are believed to be composed of building blocks conjectured in the
standard model.

The GSW $SU(2)_W \times U(1)$ electroweak theory invokes the so-called "Higgs
mechanism", in which the physical vacuum, or the true ground state, differs
from the trivial vacuum [where expectation values of all fields vanish
identically].  This theory predicts, among other things, the existence of a
Higgs particle, which has finally been observed in 2012.  Historially, many other
important predictions of the GSW electroweak theory have been substantiated
experimentally, including the existence of:  (1) neutral weak interactions, (2)
the charm quark, and (3) weak bosons $W^\pm$ and $Z^0$.  It is very likely that
any new theory which is beyond the standard model must reproduce or explain the
successes of the standard model and so will contain the standard model as a
limiting case.

Up to the moment of writing, building blocks of matter, as conjectured in the
standard model, appear to be structureless [or, less precisely, pointlike] at
the highest energy scale (or the smallest distance scale) which we are capable
of probing.  Qualitatively speaking, quarks and leptons can be described by
Dirac equations of some sort while mediators of fundamental interactions
(spin-1 particles) and the Higgs particle (spin-0 particle) are described,
respectively, by gauge field theories and a generalized Klein-Gordon equation.
It is clear that the presentation of relativistic quantum mechanics in Part A
is most relevant in the subatomic world.  Indeed, it is not clear at all
whether a Dirac equation will ever be relevant in the description of a
composite spin-${1\over 2}$ system such as a proton or a neutron.

\bigskip

\section{The Standard Model of the 21st Century}

In the year of 2012, the Standard Higgs particle is finally discovered
at CERN, Geneva, Switzerland. Subsequently in 2013, the Nobel prize
was given to Englert and Higgs for their realization of the Higgs mechanism
(or, the BEH mechanism).\footnote{\it The Universe, Vol. 1, No. 4, the Nobel
Issue. \smallskip}

There are two important, and basic, implications involved here. After searching
for the Higgs or point-like scalar particles for forty years (i.e. almost half
a century), the only thing found is the Standard-Model
Higgs particle, which represents some constrained existence of the scalar particles
(as described by the Klein-Gordon equation). The constraint for the real world, which
is difficult to spell out, amounts to the "minimum Higgs hypothesis".\footnote{\it
Thus, we put off or delete the previous presentation of the Klein-Gordon equation,
the old Chapter Two.\smallskip}

The other important implication is that the story should end at the
proper extended Standard Model - particularly, why there are three
generations? Basically, the changes among the three generations
already happen in neutrino oscillations \cite{HwangYan}. This pushed
one of us to propose the real Standard Model \cite{Hwang417}.
The Lamb's joke regarding why there is the second electron (the muon)
is indeed there. This indicates that the family idea could
be viewed as another family gauge theory \cite{Family}.

Therefore, we would like to add the 3rd pillar to the 20th-century
modern physics so that we have the 21st-century physics -- all building
blocks of matter are point-like particles and all the spin-1/2 particles
are point-like Dirac particles; referred to as "Dirac similarity principle"
to reckoned Dirac's invention of the Dirac electron, no size description
of the Dirac electron \cite{Hwang3}. It seems that all spin-1/2 building blocks of matter,
including charged or neutral leptons and quarks of various kinds, follow the route
of the Dirac electron.

Maybe we could add another one pillar - the fourth pillar of
the 21st Century. Our world seems to know the handedness of these
point-like Dirac particles -  e.g., it treats the left-handed electron
very differently from the right-handed electron. So, the basic units
of matter \cite{definition}, which treat the right-handed Dirac
particles differently from the left-handed Dirac particles, are
more appropriate than the so-called "building blocks of matter".

\bigskip

\section{The Origin of Mass}

\parindent=12pt

Suppose that, before the spontaneous symmetry breaking (SSB),
the Standard Model does not contain any parameter that is pertaining
to "mass", but, after SSB, all particles in the Standard Model acquire
the mass terms as it should --- we call it "the origin of
mass" \cite{Origin}.

At high enough temperature such at the early Universe, all the mass
terms are negligible and so set the stage of the mass generation - as
before the mechanism for the origin of mass turns on.

Originally, all the particles, fermions and bosons, are massless.
According to the origin of mass, something "ignites" the SSB in the
pure family Higgs sector; the SSB in the electroweak sector as well
as in the mixed family sector all came out as induced as a result.
The Standard-Model Higgs mass would be related to its (electroweak)
vacuum-expectation-value (VEV) by a simple factor of two. (Either
refers to the articles in "The Universe", or consult with Chapter 14
or later.)

We should remind ourselves of the fact that the generalized
Higgs mechanism is "ignited" in the purely family sector
$\Phi(3,1)$, but {\it not} in the electroweak sector
$\Phi(1,2)$ (like in the minimal Standard Model). We
believe that there is only one "ignition" point, though
in principle there could be more than one point.

Owing to the elegance of the origin of mass, we should
settle on the thinking and thus move on by treating the mass as
indicated.

This is one basic question of physics, which we should
try to answer whenever we can (since we learnt "general physics").
We think that the developments of the Standard Model
in particle physics are at the point close to actually
answering this basic question. Keep in tune on this question
when we are in progresses on this textbook.

\bigskip

\section{The Origin of Fields (Point-like Particles)}

In our world, i.e., the quantum 4-dimensional Minkowski space-time with
the force-fields gauge group $SU_c(3) \times SU_L(2) \times U(1)
\times SU_f(3)$ built-in from the outset, we realize that only the
Higgs fields $\Phi(1,2)$ (the Standard Higgs), $\Phi(3,2)$ (the
mixed family Higgs), and $\Phi(3,1)$ exist and only they could exist.
Here the first label is for $SU_f(3)$ while the second for $SU_L(2)$.

The quark world, having the (123) symmetry, is acceptable by
our world while the lepton world, having another (123) symmetry,
is also acceptable by our world. This (123) symmetry, or
nontrivial under $SU_c(3) \times SU_L(2) \times U(1)$, makes
the quark world asymptotically free and free of Landau's
ghost.

The magic comes from that, for complex scalar fields $\phi(x)$,
the interaction terms $\lambda (\phi^\dagger\phi)^2$ are
repulsive and renormalizable but that, for two "related"
complex scalar fields, the "attractive" interaction
$ -2\lambda (\phi_a^\dagger \phi_b)\cdot (\phi_b^\dagger
\phi_a)$ could overcome the repulsiveness, to rewrite
the story. The value of the universal $\lambda$
is determined by the 4-dimensional nature on the
Minkowski space-time, {\it not} by the individual
field itself. For this magic, we write the story in
"The Universe". Altogether, we call it the origin
of fields (point-like particles) \cite{Fields}.

So, in view of the repulsive nature of the self-interaction
$\lambda (\phi^\dagger \phi)^2$, the complex scalar field
$\phi$ cannot exist by itself. The three related complex scalar
fields $\Phi(1,2)$, $\Phi(3,2)$, and $\Phi(3,1)$ can co-exist,
to couple with the force-fields gauge fields (and to make them
massive, if necessary), thus making the whole story.

We believe that the description of point-like (Dirac or
Higgs) particles in terms of fields may be very
fundamental, indeed. This point is something against
what some of us venture out for alternate options (such
as superstrings).

\bigskip

\section{Prelude to Relativistic Quantum Mechanics and Quantum Fields}

When we treat a hydrogen atom, as the example, we have
to separate the center-of-mass (CM) motion from the internal
(relative) motion, exhibiting the nontrivial characteristics
associated with the internal motion. The atomic size of the
anstron ($10^{-8}\, cm$) scale is thus showing up. The
Einstein's relativity principle in the Minkowski
space-time is mostly associated with the CM motion.

QCD makes the systems to much smaller in size, about the
Fermi ($10^{-13}\, cm$) scale. The other size making, rather
than the atomic size making, can also be understood in terms
of the CM versus internal separation. What is in between
is the quantum 4-dimensional Minkowski space-time, which
allows us to chop the space-time at different scales.

Quantum fields are to be used in, e.g., electrodynamics
(in Chapters 9 and 10) or the Standard Model (after
Chapter 12 and more). Quantum fields manifest in our
real world.

Hopefully, the Standard Model which we are talking about
is not only consistent but also complete. Then, we have
something in common jointly in mathematics and physics.
The commonness is shared by jointly in mathematics in
our thinkings (and in the symbolic logic) and in
(physical) observations in our real world.

\bigskip

\section*{Appendix: Natural Units}

The brief summary in the preceding section concerning the building blocks of
matter should have made it clear that a knowledge of relativistic quantum
mechanics and quantum fields is most relevant in the area of elementary
particle and nuclear physics.  Since the kinetic energy of a particle under
investigation is often more important than its rest mass in a typical particle
or nuclear physics problem, it is convenient to measure a given velocity in
units of the light velocity $c$.\footnote{\it Note that we do not quote in
this book all the significant figures of many
basic constants which have been measured with great precision.\smallskip}
\begin{equation}
{c=2.9979 \times 10^{10} cm/sec.}
\end{equation}

\parindent=0pt
Similarly, it is convenient to express the action in units of $\hbar c$:
\medskip
\begin{equation}
\hbar c =197.33 \, MeV-fm,
\end{equation}

\parindent=0pt
where ${1 fm \equiv 10^{-13} cm}$.  The remaining units can be chosen as
powers of $MeV$
(or $GeV$) or as powers of $fm$ or as powers of $seconds$.  For instance,
a delta
width of $110 \, MeV$ corresponds to
\begin{eqnarray}
& 110 MeV\cdot {1\over \hbar c} \cdot 10^{13} fm/cm\cdot c \nonumber\\
= & 110 MeV \cdot (197.33 MeV-fm)^{-1} \cdot 10^{13} fm/cm \cdot 2.9979 \times
10^{10} cm/sec\nonumber\\
= & \{  5.98\times10^{-24} sec \}^{-1} ,
\end{eqnarray}

\parindent=0pt
or to a very short lifetime of $5.98 \times 10^{-24} sec$.  The system in which
quantities are measured in units of c and $\hbar c$ is referred to as "natural
units".  Customarily, one writes

\begin{equation}
\hbar = c = 1 .
\end{equation}

\parindent=12pt
For problems of present-day particle and nuclear physics, adoption of natural
units leads to equations which look simpler than those obtained in ordinary
units, although the physical content remains the same.  Natural units are used
in the present volume.  The situation may be considered as different from
ordinary quantum mechanics where the role played by $\hbar$ should always be
emphasized.

Finally, we note that the metric used by us (Wu and Hwang) is the same as that
by W. Pauli, T. D. Lee, H. Primakoff, and others. R. P. Feynman used both
notations and vice versa. This is the so-called "old-fashioned" notations.
This choice involves the hermitian gamma matrices on the Dirac equation.
In fact, one of us (Hwang) began on the notations by Bogoliubov but, later,
switched to (the mentor) Henry Primakoff - and fixed from there on. The choice
of notations does not affect the physics which we are learning. When there is a
need to differentiate the upper and lower indices (such as equations in general
relativity), we will make it clear when necessary - mainly in a couple of
appendices throughout this textbook.

\begin{equation}
g_{\mu\nu} = \delta_{\mu\nu}; \ \ \  \mu, \nu \in (1, 2, 3, 4).
\end{equation}

\parindent=0pt
For instance, the four-momentum of a particle is specified by

\begin{equation}
p_\mu = (\rm {\bf p}, iE),\quad with\quad p_4 = ip_0 = iE .
\end{equation}

\parindent=0pt
The inner product of two four-vectors $A_\mu$ and $B_\mu$ is given by\footnote{\it As a convention, we
use a bold-faced Roman letter or a Greek letter to denote a three-vector in
configuration space.  The arrow is reserved only for a vector in isospin space
or in other internal space.\smallskip}

\begin{equation}
A\cdot B = A_\mu B_\mu = \rm{\bf A}\cdot \rm{\bf B} - A_0 B_0 .
\end{equation}

\parindent=0pt
The only complication arises when one tries to take the complex conjugate of a
complex four-vector.  For instance, we have

\begin{equation}
\xi^*_\mu = (\xi^*, i\xi^*_0) ,
\end{equation}

\parindent=0pt
for a complex polarization four-vector  $\xi_\mu = (\xi, i\xi_0)$, so
that $\xi^*\cdot p = \xi^*\cdot \rm{\bf p} - \xi^*_0 p_0$. We shall write
out the expressions explicitly in the case that some confusion may arise.

\parindent=12pt

As the other final remark, we note that\footnote{\it To benefit readers
in the area of atomic physics, certain equations (up to Chapter 8) are
written in ordinary (m.k.s.a.) units as footnotes with primed equation
numbers. \smallskip} equations and results in the previous volume entitled
"Quantum Mechanics"\footnote{\it Henceforth referred to as "Vol. I", authored
by one of us, T.-Y. Wu, and published by World Scientific, 1986.}
will be used occasionally in this book. As such referencing
occurs, we shall use notations such as "(VIII-58), Vol. I" or "p. 12, Vol. I"
where "(VIII-58)" is the equation number and "p.12" indicates the page number.


\bigskip

\begin{thebibliography}{99}

\bibitem{Book} Ta-You Wu and W-Y. Pauchy Hwang, "Relativistic Quantum
Mechanics and Quantum Fields" (World Scientific 1991); to be updated into
the 21st Edition (the second edition).

\bibitem{Snyder} H. S. Snyder, Phys. Rev. {\bf 71}, 38 (1947).

\bibitem{PDG} Particle Data Group, "Review of Particle Physics",
J. Phys. G: Nucl. Part. Phys. {\bf 37}, 1 (2010); and its biennual
publications.

\bibitem{HwangYan} W-Y. Pauchy Hwang and Tung-Mow Yan, The Universe,
Vol.1, No.1, 5 (2013); arXiv:1212.4944 [hep-ph] 20 Dec 2012.

\bibitem{Hwang417} W-Y. Pauchy Hwang, arXiv:1304.4705 [hep-ph] 17 April
2013.

\bibitem{Family} W-Y. Pauchy Hwang, Nucl. Phys. {\bf A844}, 40c (2010);
{\it ibid.}, International J. Mod. Phys. {\bf A24}, 3366 (2009);
{\it ibid.}, Intern. J. Mod. Phys. Conf. Series {\bf 1}, 5
(2011); {\it ibid.}, American Institute of
Physics 978-0-7354-0687-2/09, pp. 25-30 (2009).

\bibitem{Hwang3} W-Y. P. Hwang, arXiv:11070156v1 (hep-ph, 1 Jul 2011), Plenary
talk given at the 10th International Conference on Low Energy Antiproton Physics
(Vancouver, Canada, April 27 - May 1, 2011). "Dirac Similarity Principle" and
"Minimum Higgs Hypothesis" were first mentioned there.

\bibitem{definition} W-Y. Pauchy Hwang, arXiv:1409.6077v1 [hep-ph] 22
Sep 2014.

\bibitem{Origin} W-Y. Pauchy Hwang, The Universe, {\bf 2-2}, 47 (2014).

\bibitem{fields} W-Y. Pauchy Hwang, The Universe, {\bf 3-1}, 3 (2015).

\end{thebibliography}
\end{document}